\documentclass{aa501}
\usepackage{graphicx}
\usepackage[]{natbib}
\bibpunct{(}{)}{;}{a}{}{,}
\begin{document}
\title{B1422+231:The influence of mass substructure on strong lensing}


   \author{M. Brada\v{c} \inst{1}
               \and
          P. Schneider \inst{1}
		\and
	  M. Steinmetz \inst{2}
		\and
	  M. Lombardi \inst{1}
		\and
          L. J. King \inst{1,3} 
		\and
	  R. Porcas \inst{4}	
          }

   \offprints{Maru\v{s}a Brada\v{c}}
   \mail{marusa@astro.uni-bonn.de}	

   \institute{Institut f\"{u}r Astrophysik und Extraterrestrische 
              Forschung, Auf dem H\"ugel 71, D-53121 Bonn, Germany
    \and Steward Observatory, 933 North Cherry Avenue, Tucson, AZ
   85721, USA
    \and Max Planck Institut f\"{u}r Astrophysik, Karl-Schwarzschild
   Str. 1, D-85748 Garching bei M\"{u}nchen, Germany
    \and Max-Planck-Institut f\"{u}r Radioastronomie, Auf dem
   H\"{u}gel 69, D-53121 Bonn, Germany
}		
\date{Submitted to A\&A, December 3, 2001}	
%
%

\abstract{In this work we investigate the gravitationally lensed
system B1422+231. 
High-quality VLBI image positions, fluxes and shapes as well as 
an optical HST lens galaxy position are used. 
First, two simple and smooth models for the lens galaxy are applied to 
fit observed image positions and fluxes; no even remotely
acceptable model was found. Such models also do not accurately 
reproduce
the image shapes. 
In order to fit
the data successfully, mass substructure has to be added to the
lens, and its level is
estimated.
To explore expectations about the level of substructure in galaxies
and its influence on strong lensing,
N-body simulation results of a model galaxy are employed. 
By using the mass distribution of this model galaxy as a lens, 
synthetic data sets of different four image system configurations are
generated and simple lens models are again applied to fit them. The
difficulties in fitting these lens systems turn out to be similar to
the case of some real gravitationally lensed systems, thus possibly 
providing
evidence for the presence and strong influence of substructure in the 
primary lens galaxy. 
}

\maketitle
%
%

\def\g{\gamma_1}
\def\gg{\gamma_2}
\def\eck#1{\left\lbrack #1 \right\rbrack}
\def\eckk#1{\bigl[ #1 \bigr]}
\def\rund#1{\left( #1 \right)}
\def\abs#1{\left\vert #1 \right\vert}
\def\wave#1{\left\lbrace #1 \right\rbrace}
\def\ave#1{\left\langle #1 \right\rangle}

\def\vc#1{%
  \if\alpha#1\mathchoice
    {\mbox{\boldmath$\displaystyle#1$}}%
    {\mbox{\boldmath$\textstyle#1$}}%
    {\mbox{\boldmath$\scriptstyle#1$}}%
    {\mbox{\boldmath$\scriptscriptstyle#1$}}%
  \else
    \textbf{\textit{#1}}%
  \fi}
\newcommand{\araa}[0]{ARAA}
\newcommand{\apj}[0]{ApJ}
\newcommand{\apjl}[0]{ApJ}
\newcommand{\aj}[0]{AJ}
\newcommand{\aap}[0]{A\&A}
\newcommand{\aaps}[0]{A\&ASS}
\newcommand{\apjs}[0]{ApJSS}
\newcommand{\bams}[0]{Bull. Amer. Math. Soc}
\newcommand{\mnras}[0]{MNRAS}
\newcommand{\pasp}[0]{PASP}
\newcommand{\nat}[0]{Nature}
\newcommand{\physrep}[0]{Phys.~Rep.}
\newcommand{\newa}[0]{NewA}

%
%
\section{Introduction \label{sc:intro}}
Gravitational lens systems with multiply imaged quasars are an
excellent tool for studying the properties of distant galaxies. In
particular, they provide the most accurate mass measures for the
lensing galaxy. Besides the mass profiles, one can also gain information
about evolution \citep{ko00} and extinction laws \citep{fa99}. Strong 
lensing is also a very promising and robust tool to
measure the Hubble constant \citep{re64}. The success of 
this method, however, depends strongly on how well the mass model is
constrained.

It turns out that image positions can be fit
quite accurately with simple, smooth elliptical models
\citep{ke97}. Since the number of observational constraints 
from image positions is small, one wants to include
the flux information. The optical fluxes, however, should
not be used, as they might be affected by microlensing and/or dust
obscuration \citep{ch79}. Radio fluxes, on the other hand, can provide 
further constraints in lens modelling.

Fitting the fluxes very accurately turns out to be
difficult in many lens systems. Models for  MG0414+0534
\citep{fa97,ke97}, PG1115+080 \citep{ke97b} and B1422+231
(e.g. \citealt{ko94b}) all show the same failure, namely 
the observed flux ratios are very different from what one would expect for
the image configurations from a smooth model. In particular, the
gravitational lens system B1422+231 was explored in detail, and as
was first mentioned by \citet{ma98}, mass substructure in the lens
galaxy might provide an
explanation for the failures in flux modelling.

A question arises as to whether there is something special about
these quadruple systems, 
or does the discrepancy simply arise
due to the fact that our smooth models are oversimplified? 
In other words, we are asking how ``well'' 
(for the purpose of strong lensing) can the smooth models used to fit
the data represent a real galaxy? N-body simulation data can provide a
realistic description of a galaxy mass distribution, thus giving a
possibility to probe its effect on strong lensing.

In this paper, we will study the influence of the
substructure on the lens system B1422+231, using VLBI radio measurements of the system 
by \citet{pa99}. In Sec.~\ref{sc:1422} we give a description of the
lens system and data used. The method is outlined in
Sect.~\ref{sc:lm} and the results on
fitting the system with smooth mass model are presented. In
Sect.~\ref{sc:mss} the model accounting for the substructure is presented
and in Sect.~\ref{sc:ell} the deconvolved image shape information is
added to the fit. 
Sect.~\ref{sc:nbody} gives the description of the
method used to investigate lensing by an N-body
simulated galaxy. We describe how we obtained synthetic data of four
image systems, and the results of fitting such systems with lens models
are presented. Finally we draw some conclusions in Sect.~\ref{sc:concl}.

This work is an abbreviated version of \citet{diploma}. 
In the course of writing this paper, several related
papers on substructure of lens galaxies have been submitted
\citep{me01,ch01,da01,me01b,ke01}. In \citet{me01} and
\citet{ch01}, the 
authors use semi-analytical descriptions to account for 
mass clumps typical for N-body simulations in a statistical fashion. 
In addition, \citet{ch01}
tests his prediction on two known systems with four images.
\citet{da01}, \citet{me01b}, and \citet{ke01} predict 
the properties of
substructure needed to constrain lens systems and compare their
results with the CDM predictions. In addition, \citet{ke01}
accounts for the difference between radio and optical fluxes by
investigating radio and optical sources of different sizes. The
present work is
differs from the afore mentioned in that slightly different models 
for the lens in
B1422+231 system
are used and that we also include image shape constraints in the fit. 
Further, we investigate four image systems generated by
using a CDM N-body simulated galaxy, rather than
an analytic approximation for the statistics of mass-substructure.

\section{\label{sc:1422}The mystery of B1422+231}

The gravitational lens system B1422+231 was discovered in the course of 
the JVAS survey (Jordell Bank -- VLA Astrometic Survey) by
\citet{pa92}. It consists of four image components. The three brightest 
images A, B, and C (as designated by \citealt{pa92}) are fairly 
collinear. The radio flux ratio between images A and B is approximately
0.9, while image C is fainter (flux ratio C to B is approximately
0.5). Image D is further away and is much fainter than the other images (with
flux ratio D:B of 0.03). The most recent available radio data for the 
image positions and
fluxes were obtained from
the polarisation observations made at $8.4 \; {\rm GHz}$ using the
VLBA and the 100m telescope at Effelsberg from \citet{pa99} and are
listed in Table~\ref{tab:42.1}.
For each of the components, the authors measured positions
(relative to the image B) and fluxes as well as the deconvolved image
shapes.   Here and through the paper we
are using a notation where  $(\theta_1,\theta_2)$ are the angular
coordinates in the
lens plane and $(\beta_1,\beta_2)$ in the source plane. 
$\theta_1$ and $\beta_1$ increase in the negative RA direction.

The radio source of this lens system is 
associated with a $15.5 \; {\rm mag}$ quasar at a redshift of $3.62$ 
\citep{pa92}. The 
lensing galaxy has been observed in the optical; its redshift has been 
determined 
to be $0.338$ and its position relative to image B has been measured 
\citep{im96}. The main lens 
galaxy is a member of a compact group with a median
projected radius of $35 \; h^{-1} \; {\rm kpc}$ and velocity dispersion 
of $\sim 550 \; {\rm km \: s^{-1} }$ \citep{ku97}.

Several groups have tried to model B1422+231
\citep{ho94,ko94b,ke97,ma98} and all of them have experienced
difficulties in fitting it. As we used data with even more precise image
positions one might expect that it would become even harder to model
the system. However, as already pointed out by some authors, the difficulties 
do not lie in fitting the image positions but rather in the
flux ratios.

It turns out that simple, smooth models fail to reproduce the radio as
well as the optical flux ratios of the system. While the
mismatch in optical data might still be due to the microlensing 
and/or dust obscurations, this can probably not explain why such models  are
not successful when fitting radio flux ratios.  \citet{ma98}
have proposed a lens model that accounts
for the substructure in the lensing galaxy. They concluded that one
needs a surface mass density  perturbation of the order of 1 \%
of the critical surface mass density in order to change the flux
ratios to the observed values.

In order to succeed in fitting the image flux
ratios, one needs to consider more sophisticated models. However, such
models also require the use of additional parameters. Therefore it is very
difficult to  ensure a constrained model that accounts for the
substructure using as constraints only image
positions, flux ratios, and the galaxy position.  
For this reason we also included  the axis ratios 
and the
orientation angles of the deconvolved images as additional constraints. 
They were obtained from \citet{pa99} 
and are given in Table~\ref{tab:42.2b}. Listed are
the absolute values of ellipticities  
and the orientation angles of the fitted elliptical
Gaussians together with uncertainties.

\begin{table}[b!]
\caption{Image positions with respect to
image B  and radio fluxes taken from \citet{pa99},where the 
uncertainties of the image positions were estimated to be 
$1/20$th of the image size in the corresponding direction (note that
these directions do not coincide with $\theta_1$ and $\theta_2$
directions). For
simplicity we set the errors to be the same in both
directions and assign
the value of $0.05 \; {\rm mas}$ for all images. $F^{i,\rm obs}$ is
the total radio flux density. The
position of the galaxy (designated by G, measured in the optical) was 
taken from \citet{im96}. The uncertainty in measuring the galaxy
position is much larger than any potential misalignment between the
optical and radio reference frames.}
\begin{center}
\begin{tabular}{c r r r r r}
\hline
Image & $\Delta \theta_{1}$ & $\Delta \theta_{2}$ & $\sigma_{\rm RA,Dec}$
&$F^{i,\rm obs}$ & $\sigma_{i,\rm flux}$  \\
 & in mas &  in mas &  in mas & mJy &  mJy \\
\hline
A & $-389.25$ & $319.98$ & $0.05$ & $152$ & $2$ \\
B & $0.0$ & $0.0$ & & $164$ & $2$  \\
C & $333.88$ & $-747.71$ & $0.05$ & $81$ & $1$ \\
D & $-950.65$ & $-802.15$ & $0.05$  & $5$ & $0.5$ \\
G & $-717$ &$-640$      & $8$& & \\ 
\hline
\end{tabular}
\end{center}
\label{tab:42.1}
\end{table} 

\begin{table}[b!]
\caption{The absolute ellipticity 
$\abs{\epsilon_{\rm i}}$ -- defined in Eq.~(\ref{eq:r.2}) -- and the
position angle $\varphi_{i}$ (measured w.r.t. the $\theta_1$-axis) of the deconvolved image $i$ 
of the fitted elliptical
Gaussians calculated from \citet{pa99} data. The uncertainty of the
absolute ellipticity $\sigma_{\abs{\epsilon},{i}}$
is determined by taking the uncertainty on the major and minor axis to
be a tenth of the beam size, which corresponds to $0.1 \; {\rm mas}$
and we considered them to be uncorrelated 
}
\begin{center}
\begin{tabular}{c r r r r }
\hline
Image & $\abs{\epsilon_{i}}$ & $\varphi_{i}$
&$\sigma_{\abs{\epsilon},{i}}$ &$\sigma_{\varphi,i}$  \\
\hline
A &0.70&  $143^{\circ}$&0.07& $5^{\circ}$\\
B &0.80&  $133^{\circ}$&0.07& $5^{\circ}$\\
C &0.55&  $106^{\circ}$&0.09& $5^{\circ}$\\ 
D &0.20&  $33^{\circ}$&0.10& $20^{\circ}$\\
\hline
\end{tabular}
\end{center}
\label{tab:42.2b}
\end{table}

From the configuration and fluxes of the four images 
we can gain some qualitative constraints on the lens model.  
Image D is much fainter than the rest which can mean that it is either
highly demagnified or that the other three images are highly
magnified. Because the position of the primary lens galaxy is known, 
the possibility of image D being highly demagnified is ruled
out. Namely, image D
does not lie at a position of high surface mass
density and thus high demagnification. Images A, B, and C are therefore
highly magnified. In order to get three highly magnified images with
a fairly collinear configuration, the source has to lie close to (and
inside) a cusp (e.g. \citealt{sc92} and references therein).

For a source position sufficiently close to and inside a cusp 
there exists a relation which states that the sum of the fluxes of the
outer two magnified images (in our case images A and C) is the same 
as the flux of the middle
image -- B (e.g. \citealt{ma92,we92}). This rule is strongly violated in the
lens 
system B1422+231. 
Actually, one can get a deviation from this relation if
the source is away from the cusp or if there is substructure in the
system (i.e. fluctuations on a scale smaller than the separations between
A, B, and C). Also the directions of elongation of the images
indicate 
that the source is
indeed close to a cusp and thus argue in favour of
substructure in the system \citep{ma98}.

%
%

\section{Lens modelling \label{sc:lm}}
First we considered two standard gravitational lens models since
their application to the \citet{pa99} data has not yet been discussed in
the literature.
\begin{table}[b!]
\caption{The summary of parameters used for the lens galaxy mass models.}
\begin{center}
\begin{tabular}{l c l }
\hline
Model& Par.& Description\\
\hline
SIE+SH& $\vc \theta_{\rm lens}$& lens position\\
 & $\sigma_{\rm v}$& Line-of-sight velocity dispersion\\
 &$\epsilon$& Absolute ellipticity\\
 & $\phi$& Position angle of ellipticity\\
 & $\gamma_1^{\rm ext},\gamma_2^{\rm ext}$& External shear
components\\
NIE+SH& $\theta_{\rm c}$& Add. to SIE+SH, core radius \\
\hline
\end{tabular}
\end{center}
\label{tab:para}
\end{table}

The standard approach to model a strong lens system is to define the
goodness-of-fit function $\chi^2$, a measure of the deviation of
the predicted and observed image properties. We perform the 
minimisation in the image plane. The
$\chi^2$ for the positions of the images reads
\begin{equation}
\chi^2_{\rm pos}=\sum_{i=1,3,4}\frac{\abs{\Delta {\vc \theta}^{i,\rm
\:obs}- \Delta \vc \theta^{
i, \rm \:mod}}^2}{\sigma_{i, \rm \:pos}^2} \;,
\label{eq:lm.7}
\end{equation}
where $\Delta \vc \theta^{i,{\rm \:obs}}$ and $\Delta \vc \theta^{i,\:{\rm
 mod}}$ 
are the observed and  the modelled position
of the $i$-th image relative to a chosen origin 
(in our case image B, denoted by $i=2$), respectively. Note that the
 sum extends only over the images A, C, and D.

To be more precise,  $\Delta \vc \theta^{i,{\rm \:mod}}$ is obtained as
\[
\Delta \vc \theta^{i,{\rm \:mod}}= \vc \theta^{i,{\rm \:mod}}-\vc \theta^{2,{\rm
\:mod}}\; ,
\]  
where the vectors $\vc \theta^{i,{\rm \:mod}}$ and  $\vc \theta^{2,{\rm \:mod}}$
are measured with respect to an independent reference point. The same
is true for  $\Delta \vc \theta^{i,{\rm \:obs}}$,
\[
\Delta \vc \theta^{i,{\rm \:obs}}= \vc \theta^{i,{\rm \:obs}}-\vc \theta^{2,{\rm
\:obs}}\; ,
\] 
which is the observed image position as listed in
Table~\ref{tab:42.2b}. 
Therefore when the image positions of A, C, and D w.r.t. the image B are 
measured, they all contain the uncertainty of the origin (image B). 
Thus, the uncertainties of the measurements of each of the image
positions are correlated and 
we are forced to calculate the $\chi^2_{\rm pos}$ as in
(\ref{eq:lm.7}). We therefore denote $\sigma_{i, \rm \:pos}$ to be  
the uncertainty of the measured relative image position.

Similarly, we define 
for the galaxy position
\begin{equation}
\chi^2_{\rm galaxy}=\frac{\abs{\Delta \vc \theta_{\rm lens}^{\rm
obs}- \Delta \vc \theta_{\rm lens}^{\rm mod}}^2}{\sigma_{\rm gal, \:pos}^2} \;,
\label{eq:lm.8}
\end{equation}
with $\sigma_{\rm gal, \:pos}$ being the uncertainty of the relative 
lens position. The contribution to the $\chi^2$-function from the 
fluxes is given by
\begin{equation}
\chi^2_{\rm flux}=\sum_{i=1}^{4}\frac{\rund{F^{
i,\rm \:obs}-F^{i, \rm \:mod}}^2}{\sigma_{i,\rm flux}^2} \;,
\label{eq:lm.5}
\end{equation}
where $F^{i,\: \rm mod}$ denotes the flux of the $i$-th 
image obtained from the model and
$F^{i, \: \rm obs}$ is the observed flux.

Further we can use deconvolved image shapes as constraints of 
the model. We
work in terms of the complex ellipticity, 
\[
\epsilon_{i}:=\abs{\epsilon_i} {\rm e}^{2{\rm i}\varphi_i}\;.
\]
Each image is thus described by
the absolute value of ellipticity $\abs{\epsilon_{i}}$ and the 
corresponding position
angle $\varphi_{i}$ (measured w.r.t. $\theta_1$-axis). 
From the given data the absolute value of ellipticity is calculated as
\begin{equation}
\abs{\epsilon_{i}}=\frac{a_{i}-b_{i}}{a_{i}+b_{i}}\; ,
\label{eq:r.2}
\end{equation}
where $a_i$ and $b_i$ are major and minor semi axes of the
fitted elliptical Gaussian. Two additional terms,
\begin{equation}
\chi^2_{\epsilon}=\sum_{i=1}^{4}
\frac{\rund{\abs{\epsilon_{i}^{\rm obs}}-\abs{\epsilon_{i}^{\rm
mod}}}^2}{\sigma_{\abs{\epsilon},i}^2}\; , 
\label{eq:r.3a}
\end{equation}
and
\begin{equation}
\chi^2_{\varphi}=\sum_{i=1}^{4}
\frac{\rund{\varphi_{i}^{\rm obs}-\varphi_{i}^{\rm
mod}}^2}{\sigma_{\varphi,i}^2}\;.
\label{eq:r.3b}
\end{equation}
have to be added to the $\chi^2$-function. For simplicity we assumed the
errors on ellipticity and position angle to be uncorrelated. The
$\chi^2$ function we want to minimise is simply given by the sum 
\begin{equation}
\chi^2_{\rm tot}=\chi^2_{\rm pos}+\chi^2_{\rm galaxy}+\chi^2_{\rm
flux} +\chi^2_{\epsilon}+\chi^2_{\varphi}\; .
\label{eq:lm.6}
\end{equation}

In order to obtain the values of  $\Delta \vc \theta^{i,\rm \:mod}$ from
the lens equation (see e.g. \citealt{sc92}), we used the Numerical
Recipes {\tt MNEWT} routine
from \citet{numrec} for solving a set of two non-linear equations. The
$\chi^2$-function was then minimised with respect to the model
parameters using {\tt POWELL}, a multi-dimensional minimisation routine, also
from \citet{numrec}.

One of the consideration one faces when doing lens modelling is 
the uniqueness of the solution. Minimisation
methods such as {\tt POWELL} do not give an answer as to whether a
minimum found is a local or global one. One can partly solve this by
using simulated annealing routines, such as {\tt AMOEBSA} from
\citet{numrec}. However, these routines are much more CPU-time consuming,
and they are also
not completely foolproof in finding an absolute minimum.

As an additional check, to minimise the chance of obtaining a secondary
minimum as a result, we have run  
{\tt POWELL} for several different
initial conditions. Unless the global minimum is extremely narrow
in parameter space (which is rather unlikely) we can be confident of
obtaining the global minimum.

\section{Modelling B1422+231 with smooth models \label{sc:rSIE+SH}}
We used a singular isothermal ellipsoid with
external shear from
\citet{ko94b} (hereafter SIE+SH) and a non-singular isothermal ellipsoid
model with external shear (NIE+SH) from \citet{ke98} 
to fit the image positions and fluxes of B1422+231. The
explanation of the model parameters for the models we used are given in
Table~\ref{tab:para}.

We have applied the fitting procedure described above to the radio data,
using image positions, fluxes, and their uncertainties from
\citet{pa99}, listed in Table~\ref{tab:42.1}.    
The optical position of the galaxy was taken from
\citet{im96}. Although the image positions are very accurate 
(of the order of $50 \;
{\mu \rm arcsec}$), we have no difficulties fitting them, and so the
$\chi^2$ contribution from the image positions drops to zero (see 
Table~\ref{tab:r.9}). However, 
as already pointed out in previous works on B1422+231,
the model completely fails in predicting the image fluxes. In particular
image A is predicted too dim (the modelled flux ratio A:B turned
out to be 0.80, much below the measured value of 0.93). We have also tried to model the system with a NIE+SH model; however, the
$\chi^2$ did not improve significantly.

Next we disregarded the flux of image A as a constraint,
trying to see whether we can get a good fit for the rest of the
images. Reducing the number of constraints by one, we are left with one
degree of freedom. The results of the fitting 
are given in Table~\ref{tab:r.9}. 
The fit is still not satisfactory, thus indicating that changes of
magnifications are needed not only for image A.

The choice of these two particular macromodels was due to their
simplicity and analytic solutions for the deflection angle. Of course
these are not the only models one can use, and one might hope to
explain the discrepant flux ratios by using another family of models.   
In order to effectively
change the flux ratios, one has to change the magnification
gradient. The simplest way of changing a gradient is to change the
power-law behaviour of the potential.

Although the success of such modelling is very unlikely (as shown by
\citealt{ma98}), we have tried to use the SIE+SH model, adding an additional
black hole with an Einstein radius fixed to $10^{-2}$ of the main galaxy
Einstein radius, centred on the lens galaxy
position. The corresponding mass of the black hole is 
$\sim 10^{8} \: {\rm M_{\odot}}$, 
in agreement with the correlation of the black hole mass with the 
velocity dispersion of its host bulge \citep{fe00}. 
Such a model
is plausible, however we found no significant improvement of the
fit (see Table~\ref{tab:r.9} for results). Also, considering the black 
hole mass as a parameter did not significantly improve the outcome.

Although other smooth models can still be investigated, it seems unlikely that
another smooth model can explain all four image fluxes
simultaneously. Even when one disregards the flux of image A, a smooth
macromodel seems to be incapable of explaining the remaining flux ratios.

\section{Models with substructure \label{sc:mss}}
In the previous section it turned out that A:B flux ratio causes 
the biggest difficulty in 
fitting the B1422+231. Since the radio and optical flux ratios are very
different, one is tempted to exclude it from the $\chi^2$ measure (see
\citealt{ma98}, \citealt{ch01}).

However, one can also try to deal with this problem in another
way. Adding a small perturber at the same angular diameter distance as
the primary lens and at approximately the same position as image A can
change the flux ratio A:B substantially. On the
other hand, calculations show \citep{ma98} that such a perturber does
not affect the positions of any of the images
appreciably. Furthermore, a small
perturber can also change the flux ratio of the other images slightly and
this might help to improve the results from the previous section.

We model the perturber as a non-singular isothermal sphere. 
This gives two additional parameters (two perturber
positions, since we keep the line-of-sight velocity dispersion and core
radius of the perturber fixed)
to the macro (SIE+SH) model. The choice of the perturber being modelled as
non-singular is due to the fact that a singular isothermal sphere (with
the same Einstein radius) is more
likely to give rise to additional (observable) images.

We are aware of the fact that the choice of this particular model
for the substructure is oversimplified in many ways. However, we are
not trying to constrain the nature of substructure in this case; which
is impossible due to the number of constraints
available. In the following we investigate whether 
the peculiarity of the fluxes can be explained by a small satellite
galaxy in the neighbourhood of one
(or two) of the images.

\subsection{Modelling B1422+231 using the substructure model
\label{sc:rSIE+SH+NIS}}
 
For modelling B1422+231 we fixed the line-of-sight velocity dispersion
of the NIS perturber to   $\sigma_{\rm
NIS}=10 \; {\rm km\,s^{-1}}$ (equivalent to an Einstein radius of
approximately $2\:{\rm mas}$) and its core radius to  
$\theta_{\rm c}=20\;  {\rm mas}$. A  
perturber with these properties does not affect the
image positions significantly; on the other hand it can substantially
change the magnification at the position of one of the images.

When fixing the core radius one has to be aware that not only an 
SIS, but also an NIS lens
with small enough core radius might give additional images.    
Therefore, an NIS  perturber should have a core radius much bigger than
the Einstein radius, in order not to produce additional
images. We have checked that indeed no
additional observable images are predicted by the model.

For this purpose we define the function $f$ on a
grid of points $\vc \theta_j$  
\begin{equation}
f=\abs{\vc \beta(\vc \theta_j) -\vc \beta_{\rm s}}^2\; ,
\label{eq:r.7}
\end{equation}
where $\vc \beta(\vc \theta_j)$ is the calculated source position 
corresponding to
the point $\vc \theta_j$. The position $\vc \beta_{\rm s}$ is defined 
as the average 
position for the
source predicted by the four observed image positions 
\begin{equation}
\vc \beta_{\rm s}=\frac{1}{4}\sum_{i=1}^4 \vc \beta \rund{\vc \theta^{i,\rm \: obs}}\;.
\label{eq:lm.2}
\end{equation} 
The function $f$ vanishes only around the
positions where images are observed, provided that the grid
resolution is high enough. In our case the grid resolution
was $0.05\: {\rm mas}$ (small compared to the Einstein radius of the
perturbing galaxy, $\theta_{\rm E, NIS}=3 \: {\rm mas}$). We indeed 
found only four such regions, and they correspond to the four observed
images. 
\begin{table}[b!]
\caption{Result of modelling B1422+231 radio positions and flux data 
with (i) an SIE model with external shear 
(ii) SIE+SH without the flux of image A as a constraint and (iii)
SIE+SH model with an additional black hole at the galaxy position
without the flux of image A as a constraint.  The black
hole Einstein radius in model (iii) was fixed to $10^{-2}$ of the main galaxy
Einstein radius. The parameters of the
best fitting model are listed in Table~\ref{tab:all}.}
\begin{center}
\begin{tabular}{c c c @{} c @{} c @{} c @{} c @{} c @{} c}
\hline
Model& $N_{\rm dof}$ & $\chi_{\rm tot}^2$&${}={}$&$\chi_{\rm pos,rel}^2$&${}+{}$&
$\chi_{\rm gal,rel}^2$&${}+{}$&$\chi_{\rm flux}^2$ \\
\hline
(i)&2& 129&${}={}$&0&${}+{}$&18&${}+{}$&111\\ 
(ii)&1& 9.8&${}={}$&0&${}+{}$&0.5&${}+{}$&9.3\\
(iii)&1& 7.5&${}={}$&0&${}+{}$&0.2&${}+{}$&7.3\\
\hline
\end{tabular}
\end{center}
\label{tab:r.9}
\end{table}

The resulting model has 12 parameters,
which leaves us 0 degrees of freedom. 
The $\chi^2$ has decreased by a factor of more than
20 compared with the SIE+SH model; however, since we have zero degrees
of freedom we expect
$\chi^2$ to vanish if the model is realistic (and if the $\chi^2$
technique is an adequate method). 
The family of models considered thus does not seem to be
adequate for the description of the galaxy in B1422+231 lens system.

We again see that the flux ratio of images A:B is
not the only problem when dealing with fluxes. Adding
a small perturber close to image A allows us to adjust the modelled
flux of A such, that it does not give any contribution to the
$\chi^2$-function. Still the remaining image fluxes are not fit
perfectly, leading to a possible conclusion that all images are
affected by the mass substructure.

\subsection{Using image shapes as constraints \label{sc:ell}}
In order to increase the number of degrees of freedom we are dealing with,
further constraints have to be added. We therefore used the
deconvolved image shapes (see Table~\ref{tab:42.2b}). 
Due to the problems in determining the image shapes we describe
later, we will use them as constraints only in this section.

The $\chi^2$ minimisation was done according to the
previous section, and the results are presented in
Table~\ref{tab:r.13}. 
\begin{table}[b!]
\caption{Result of modelling B1422+231 radio positions, flux and
in the case of (v), (vi) and (vii) image shape data.  
The models we used were an SIE model with external shear and an 
additional perturbing NIS
galaxy (iv), an SIE model with external shear (v), SIE+SH and {\it one\/} 
additional perturbing NIS
galaxy (vi), and SIE+SH with {\it two\/} perturbing galaxies (vii).  
The core radius and $\sigma_{\rm NIS}$ of the perturbing
galaxy(ies) were
fixed to the values $\theta_{\rm c}=20\;  {\rm mas}$ and $\sigma_{\rm
NIS}=10 \; {\rm km /s}$ for model (iv) and $\theta_{\rm c}=1\;  {\rm mas}$ and $\sigma_{\rm
NIS}=10 \; {\rm km /s}$ for the other models (see text).
The parameters of the best fitting models are listed in Table~\ref{tab:all}.}
\begin{center}
\begin{tabular}{c c c @{} c @{} c @{} c @{} c @{} c @{} c @{} c @{} 
c @{} c @{} c}
\hline
Model &$N_{\rm dof}$&$\chi^2_{\rm tot}$&${}={}$&$\chi_{\rm pos,rel}^2$&${}+{}$
&$\chi_{\rm gal,rel}^2$&${}+{}$&$\chi_{\rm flux}^2$&${}+{}$
&$\chi^2_{\epsilon}$&${}+{}$&$\chi^2_{\varphi}$ \\ 
\hline
(iv)&0& 5.6&${}={}$&0.0&${}+{}$&0.8&${}+{}$&4.8& &\\ 
(v)&8& 140&${}={}$&0&${}+{}$&18&${}+{}$&111&${}+{}$&4&${}+{}$&7 \\
(vi)& 6& 19&${}={}$&0&${}+{}$&0&${}+{}$&6&${}+{}$&9&${}+{}$&4\\ 
(vii)&4& 15&${}={}$&0&${}+{}$&0&${}+{}$&4&${}+{}$&7&${}+{}$&4\\ 
\hline
\end{tabular}
\end{center}
\label{tab:r.13}
\end{table}   
We have decreased
the value of the core radius of the perturber in order to get higher 
magnification
gradients, since large changes in image shapes as predicted by a
smooth model was needed.
Again, no additional observable images are produced by
the perturber(s).

In the \citet{pa99} paper the uncertainties on the image shapes are
not listed. 
The image shapes are obtained by fitting Gaussian profiles to the
map, and then deconvolved using the known beam-shape. The uncertainties 
are therefore just a rough estimate, since one
can not quantitatively account for the error of such fitting. 
In fact, all the images exhibit non-Gaussian features a Gaussian model
is an oversimplification for the image shape description.

The resulting $\chi^2$ for the minimisation was 19, having 6
degrees of freedom (the probability of obtaining a value for
$\chi^2$ bigger than 19 is $0.0042$). We further try to fit the data 
with two perturbers in the system. 
If we put two equal perturbers into the system (i.e. both with fixed
$\theta_{\rm c}=1\;  {\rm mas}$ and $\sigma_{\rm NIS}=10 \; {\rm km /s}$), 
we get a $\chi^2$ of 15 (see Table~\ref{tab:r.13}) with 4 degrees of
freedom. The
probability of obtaining  a $\chi^2$ larger than that value is now
$0.0047$. Such a
reduction of $\chi^2$ is apparently not a significant improvement of the
fit (compared to the model with a single perturber).

Just for comparison, we also include the deconvolved image shapes in
the fit with SIE+SH model. The resulting $\chi^2$ is 140 for 8
degrees of freedom (the probability of obtaining a value for
$\chi^2$ larger than that is now only $10^{-28}$). We essentially 
get the same model as when  
only fitting image positions and fluxes (see Table~\ref{tab:all}). We
see that the fluxes provide much stronger constraints (their
uncertainty is  much smaller) than the image shapes.

Apparently, models with substructure yield significantly better fits
than the ones without. However, we should stress that, strictly
speaking, the appropriate
statistical  treatment can not be easily performed because the model parameters
enter the $\chi^2$ function in a non-linear way. As a result, the
$\chi^2$ function does {\it not\/} behave as a chi-square random
variable with the appropriate number of degrees of freedom. One can
see that already from the fact that the $\chi^2$ function did not
vanish when using a model with zero degrees of freedom. 
Hence, the probabilities we quoted above have to be taken with care;
still they can be used to compare the performance of different models.

It is clear that it is difficult to simultaneously stretch and rotate
the images with one (or two) perturber(s). The macro-model is not very
successful in predicting both components of the image ellipticities
and the fluxes, and
therefore corrections are needed in the case of all four images. 
We can safely assume that the inclusion of further subclumps in
the model would eventually lead to a ``perfect'' fit with the observed
data. In particular three more subclumps close to the images B, C, and
D would yield a significant improvement to the $\chi^2$.

\begin{table}[b!]
\caption{Resulting parameters from best fitting models:
(i) SIE+SH, (ii) SIE+SH, (iii) SIE+SH+BH, both without the flux of
image A as a constraint, (iv) SIE+SH+NIS, (v) SIE+SH, (vi) SIE+SH+NIS, (vii)
SIE+SH+2NIS. In models (v),
(vi) and (vii) we use the shapes of
the images as additional constraints. $\vc \theta_{\rm NIS}$ is 
the position of the
perturber(s) w.r.t. the image indicated in the superscript. 
The parameters of the
best fitting source
shape were  $\abs{\epsilon^{\rm s}}=0.14$ and $\varphi^{\rm
s}=60^{\circ}$ for the model (iii), for the model (iv) $\abs{\epsilon^{\rm s}}=0.04$, $\varphi^{\rm
s}=30^{\circ}$, and for the model (v) 
$\abs{\epsilon^{\rm s}}=0.10$, $\varphi^{\rm
s}=20^{\circ}$. The position angles are measured
w.r.t. $\theta_1$-axis. The resulting 
line-of-sight velocity dispersion $\sigma_{\rm v}$
was $190 \; {\rm km\:s^{-1}}$ in all four cases.}
\begin{center}
\begin{tabular}{c@{$\; \;$} c@{$\; \;$} c@{$\;\;$} c@{$\; \;$}
c@{$\;\;$} c@{$\;\;$} c@{$\;\;$} c@{$\;\;$} c@{}}
\hline
 & $\vc \theta_{\rm lens}$& $\epsilon_{\rm gal}$ & $\phi$ &$\gamma_1^{\rm ext}$ & $\gamma_2^{\rm ext}$&$\vc \theta_{\rm NIS}$\\
 &$({\rm mas},{\rm mas})$ &  & & &  & $({\rm mas},{\rm mas})$ \\
\hline
(i)& $ (-744,-659)$ & $ 0.19$ &   $34^{\circ}$  & $ -0.04 $ &$-0.16$ &
\\
(ii)& $ (-721,-643)$ & $ 0.14$ &   $33^{\circ}$  & $ -0.05 $ &$-0.17$ &
\\
(iii)& $ (-720,-643)$ & $ 0.15$ &   $33^{\circ}$  & $ -0.05 $ &$-0.18$ &
\\
(iv)& $ (-722,-646) $&  $ 0.13$ &$32^{\circ}$ &$-0.05$&$-0.18$
&$(41,36)^{\rm A}$\\
(v)& $ (-744,-659)$ & $ 0.19$ &   $34^{\circ}$  &$  -0.04$  &$-0.16 $& \\
(vi)&$ (-718,-643)$   & $ 0.12$ &   $32^{\circ}$  & $ -0.05$  &
       $ -0.18$ &$(53,47)^{\rm A}$\\ 
(vii)& $(-719,-641)$ & $ 0.13$ &   $32^{\circ}$  & $ -0.05$  &
       $ -0.18$ &$\begin{array}{c} (-18,7)^{\rm A}\\ (-5,4)^{\rm D}\end{array} $\\
\hline
\end{tabular}
\end{center}
\label{tab:all}
\end{table}

%
%

\section{Strong lensing by an N-body simulated galaxy \label{sc:nbody}}
A question that arises from model fitting of B1422+231 is 
whether such behaviour is seen by the N-body simulated galaxy and
therefore generic of a typical galaxy lens.
We used the cosmological N-body simulation data including 
gasdynamics and star formation of \citet{st01} for this purpose.
\begin{figure}[ht!]
\begin{center}
\includegraphics[width=0.45\textwidth]{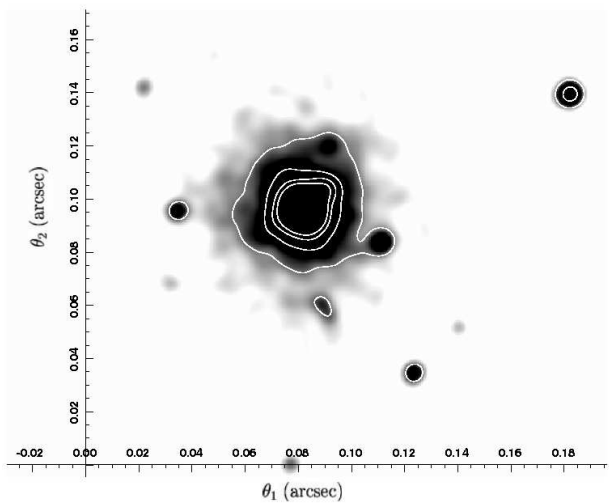}
\end{center}
\caption{The cut-out of the surface mass density map of the simulated 
galaxy. The mass distribution resulting from the cosmological N-body 
simulation (see text) was
smoothed using convolution with a Gaussian kernel characterised
by a standard deviation $\sigma \sim 0.8 \; {\rm kpc} \sim 0.2 \: {\rm
arcsec}$. 
This map was then evaluated on $2048 \: \times \: 2048$ grid points 
($\sim 160 \: \times \;160 \: {\rm kpc}$) and the surface mass
density was calculated. Test particles, used in the N-body simulation
to account for the large-scale
structure, have been removed here. The contours correspond to the values of 
$\kappa=0.8;1.6;2.4;3.2$. The dark regions represent the regions of
high $\kappa$. The units on the axes are arcseconds; one arcsecond
in the {\it lens\/} plane corresponds to approximately $4 \: {\rm kpc}$.} 
\label{fig:kappa_2k}
\end{figure}

The simulations were performed using GRAPESPH, a code that combines the
hardware N-body integrator GRAPE with the Smooth Particle Hydrodynamics
technique \citep{st96}. 
GRAPESPH is fully Lagrangian and optimally suited to
study the formation of highly non-linear systems in a cosmological
context. 
The version
used here includes the self-gravity of gas, stars, and dark matter
components, a three-dimensional treatment of the hydrodynamics of the
gas, Compton and
radiative cooling (assuming primordial abundances), the effects of a
photo-ionizing UV background, and a simple recipe for transforming gas
into stars.

The original simulated field is located at $z=0.2$ and contains
approximately 300000  particles. The simulation is contained within
a sphere of
diameter $32\; {\rm Mpc}$ which is split into a high
resolution sphere of diameter  $2.5\;{\rm Mpc}$ centred around a
galaxy and an outer low resolution shell. Gas dynamics and star
formation is restricted to the high resolution
sphere (280000 particles, 92000 of which dark matter), while the 34000 dark
matter particles of the low resolution sphere sample the large scale matter
distribution in order to appropriately reproduce the large scale tidal fields
(see \citealt{na97} and \citealt{st00} for details on this
simulation technique).

The simulation was performed in a $\Lambda$CDM cosmology
($\Omega_0=0.3$, $\Omega_{\Lambda}=0.7$, $\Omega_{\rm b}=0.019/h^2$,
$\sigma_8=0.9$).  It has a
mass resolution of $1.26 \times 10^{7} M_{\odot}$ and a spatial
resolution of $0.5 \: {\rm kpc}$. A realistic
resolution scale for an identified substructure is typically assumed 
to be $\sim 40$ particles which
corresponds to   $5 \times 10^{8} M_{\odot}$. The quoted mass
resolution holds for gas/stars. The high resolution dark matter 
particles are
about a factor of 6 ($=\Omega_{0}/\Omega_{\rm b}$) more massive.

From the original simulated field we took a cut-out map of size 
$\sim 160 \: \times \;160 \: {\rm kpc}$ that is centred on a single
galaxy. This area lies well within the high resolution sphere and is
void of any massive intruder particles from the low resolution shell. 
The resulting mass distribution was
smoothed using convolution with a Gaussian kernel characterised
by a standard deviation of $\sigma \sim 0.8 \: {\rm kpc} \sim
0.2 \: {\rm arcsec}$.\footnote{For the distance calculations 
through the paper we assumed an Einstein-de-Sitter
Universe and the Hubble constant $H_{0}=65\:{\rm km\, s^{-1}\, Mpc^{-1}}$.} 
This map was then
evaluated on $2048 \; \times \; 2048$ grid points. The final map contains
information about approximately 130000 particles with a total mass of
$3.0 \times 10^{12} \; {\rm M_{\odot}}$. The surface mass density
$\kappa$ was
calculated for every grid point.  We chose
the redshift of the source to be $z=3$. A part of the cut-out map can be seen
in  Fig.~\ref{fig:kappa_2k}.

The lens properties are calculated on a grid of $2048 \:\times
\:2048$ points. The Poisson equation 
\begin{equation}
\nabla^2 \psi(\vc \theta)=2\, \kappa(\vc \theta) \; .
\label{eq:t.16}
\end{equation}
is solved (on the grid) in Fourier space by the FFT (Fast Fourier
Transformation) method. This method is incorporated in the routine
{\tt KAPPA2STUFF} from the IMCAT software of Nick Kaiser
({\tt http://www.ifa.hawaii.edu/\~{}kaiser}) that we used.  It
takes a grid map of $\kappa(\vc \theta_j)$ as an input and returns the
values of deflection angle and complex shear (in $\vc x$-space), along
with other quantities. From these data we calculated the
Jacobi matrix for each grid point.

The simulated galaxy is a field galaxy. Therefore we add two external
shear components to the Jacobi matrix (evaluated at each grid point)
in order to make it similar to the galaxy in
B1422+231. The shear components were taken to be the same as the 
ones obtained from
the best fitting SIE+SH model
\[
\gamma_1^{\rm ext}=-0.04 \; , \quad \gamma_2^{\rm ext}=-0.16 \; .
\]
The external shear accounts for 
the effect of the neighbouring galaxies of the compact group, which are
not present in the simulation. 

\begin{figure}[ht!]
\begin{center}
\includegraphics[width=0.45\textwidth]{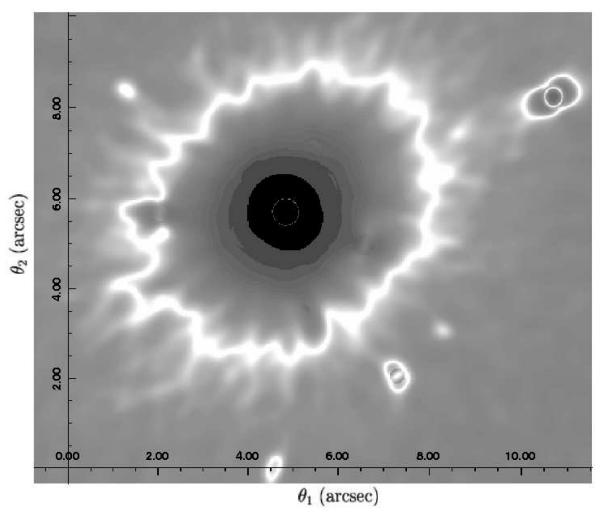}
\end{center}
\caption{The magnification map of the simulated galaxy
calculated using the {\tt KAPPA2STUFF} routine from
Nick Kaiser's software IMCAT. External shear is added in the evaluation of the
magnification map for to account for neighbouring galaxies (see
text). Lighter regions represent high magnifications. The units on the
axes are arcseconds; one arcsecond in the {\it lens\/} plane corresponds to
approximately $4 \:{\rm kpc}$.}
\label{fig:magn_map}
\end{figure}

Fig.~\ref{fig:magn_map} shows the magnification map of the surface mass
density (given in Fig.~\ref{fig:kappa_2k}) with additional external
shear. One can clearly see the outer critical curve (white curve),
while only the traces of the inner critical curve are visible (little
circle inside the black region). The
reason why we can see the outer critical curve so much better than
the inner one is the following. At the centre  
$\vc \theta_{\rm c}$ of the galaxy, the surface mass density is very high and the determinant of the
Jacobi matrix can be approximated by 
$\det A \sim \kappa(\vc \theta_{\rm c})^2$. In
our particular case, $\kappa(\vc \theta_{\rm c}) \approx 50$, and since at the
critical curve $\det A =0$, we see that the determinant has to decrease
from 2500 to 0 in a region of $0.4 \; {\rm arcsec}$. The transition is therefore
very steep and we have a very good chance to miss the maximum value of
magnification there, since the resolution is not high enough. At the
outer critical curve, the change is slower and we can clearly
see the points of high magnification.

\begin{figure}[t!]
\begin{center}
\includegraphics[width=0.45\textwidth]{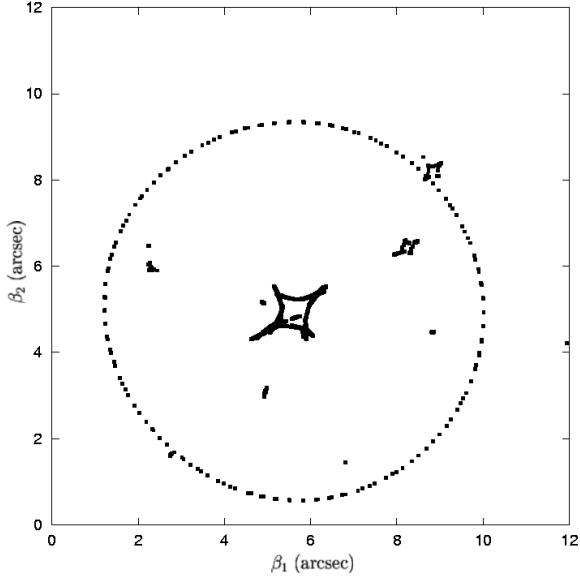}
\end{center}
\caption{The caustic obtained by mapping the points of high
magnification  onto the source plane. 
We took $\abs{\mu}>30$ for the inner caustic, for the outer
one we additionally picked the points of $\abs{\mu}>0.5$ (due to
nummerical effects -- see text) from the
central part of the magnification map. The units on the axes are
arcseconds; one arcsecond in the {\it source\/} plane corresponds to
approximately $6 \: {\rm kpc}$ for the source at $z=3$.} 
\label{fig:nb_caustic}
\end{figure}

In order to generate a similar image configuration as the one in 
B1422+231,
one considers the caustic curve. This can be done by simply
mapping the points of high magnification onto the source plane.
Such a map is presented in Fig.~\ref{fig:nb_caustic}.

\begin{figure}[t!]
\begin{center}
\includegraphics[width=0.45\textwidth]{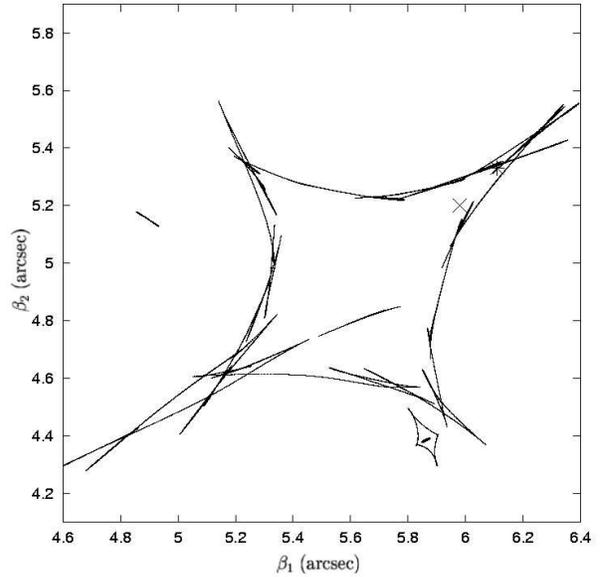}
\end{center}
\caption{The caustic curves obtained by first interpolating the
magnification map on a refined grid using bilinear interpolation
(increasing the number of points in the region of interest by a factor
of 25) and mapping the points of high
magnification $\abs{\mu}>30$ (Fig.  \ref{fig:magn_map}) to the source
plane. The units on the axes are arcseconds; one arcsecond in the 
{\it source\/} plane corresponds to
approximately $6 \: {\rm kpc}$ for the source at $z=3$. Two marks
correspond to the source position of data set 1 (cross) and 11 (star)
-- see Sect.~\ref{sc:synth}. The source positions of other
data sets are located on a line connecting them.} 
\label{fig:c_inter}
\end{figure}

On first sight the caustic
structure of the N-body simulated galaxy looks the same as e.g. the
caustic of the smooth NIE model with a small core radius. 
However, if we look only at the inner,
asteroid caustic we can see that it is not completely ``smooth''. With
the help of bilinear interpolation we recalculated the magnification map
on a refined grid (increasing the number of points being evaluated by 
$5 \: \times 5$) and the corresponding caustic for such a grid 
is shown in Fig.~\ref{fig:c_inter}. The caustic structure is much more
complicated than in the case of a smooth model; in
addition to folds and cusps we also have swallowtails formed (see
\citealt{sc92} for more details on catastrophe theory).

\subsection{Generating synthetic data \label{sc:synth}}
We select a source position $\vc \beta_{\rm s}$ such
as to lie
inside the asteroid caustic and close to the cusp, trying to
chose a position for which we would get similar flux ratios as in
the case of B1422+231. 
In total we considered 11 different
source positions (see Fig.~\ref{fig:c_inter}).

For each of them we first determine approximate image positions
using the method described in Sect.~\ref{sc:rSIE+SH+NIS}. 
In order to get exact image positions we use the root
finding method {\tt MNEWT} again, for which we need the deflection angle to be
continuous inside the region where we look for images. In our case the
deflection angle is defined only on a $2048 \times 2048$ grid. We 
perform bilinear interpolation between grid points.  
Having the image positions, we performed bilinear
interpolation of the magnification map in order to get more exact
magnification factors.

\subsection{Fitting the synthetic data \label{sc:fit}}
For the $\chi^2$-fitting method according to (\ref{eq:lm.6}) we need
to determine the uncertainty on the image positions and fluxes. Since
we use interpolation for
the {\tt MNEWT} method we do not have a real estimate for the errors. 
One can, for example, set the
errors to the same (relative) values as the uncertainties on the
observed radio positions in B1422+231. For the typical scales we are
using 
here (i.e. the distance B to D is approximately  $3.5 \;
{\rm arcsec}$) this would mean an
uncertainty of much less than a distance between two grid points. However, such a small error estimate
is not realistic; due to the finite grid we estimate 
the image position uncertainties to be the distance between two grid
points (which is a generous
estimate; we use bilinear interpolation so the uncertainty is
probably lower). The flux ratio errors
were then set to be approximately 2 \% for
images A, B and C and 5 \% for image D. These uncertainties are set
to be the same as in the case of observed radio fluxes in B1422+231. The
galaxy position error is set twice as big as that for the image positions.

The fitting procedure is performed in the same way as for the
B1422+231 data. 
Again, image B is taken as a reference and
the $\chi^2$-function is evaluated according to (\ref{eq:lm.6}). We
try to fit the positions and fluxes with SIE+SH and SIE+SH+NIS
models. The flux ratios of the 11 sets of synthetic data, together with
the results of the minimisation are presented in 
table \ref{tab:dataset}. We experience similar problems 
fitting fluxes as before; the $\chi^2$-function value is high for
all 11 data sets.

What might be surprising is the fact that we do not recover
some properties of the lensing system.  We know that the primary lens
is fairly circular (one can {\it not\/} see that from fig. \ref{fig:magn_map},
since there, external shear is already added) and we know the values of
the external shear components. What we also know a priori are the magnification
factors for the images. These values were not
recovered with high accuracy in model fitting. 
\begin{table}[b!]
\caption{The flux ratios of image A, C and D w.r.t. image B of 11 data sets with different image positions
$\vc \beta_{\rm s}$ (see also Fig.~\ref{fig:c_inter}). Listed are
also the resulting $\chi^2$ values for fitting image positions
and fluxes with SIE+SH model ($\chi^2_{1}$) and SIE+SH+NIS model 
($\chi^2_{2}$). The core radius and $\sigma_{\rm NIS}$
of the perturbing galaxy were fixed to the values 
$\theta_{\rm c}=1\;  {\rm mas}$ and $\sigma_{\rm NIS}=15 \; {\rm km /s}$. }
\begin{center}
\begin{tabular}{c r r r r r }
\hline
Data set & $F_{\rm AB}$ & $F_{\rm CB}$  & $F_{\rm
DB}$& $\chi_{1}^2$&$\chi_{2}^2$\\
\hline
1&     1.04&        0.80&        0.220& 	120&2.0	\\        
2&     1.29&        1.20&        0.193&	960&120	\\         
3&     1.43&        1.40&        0.167&	2100&190\\          
4&     1.20&        0.79&        0.142&	660&5.1	\\          
5&     1.06&        0.59&        0.116&	350&3.9	\\        
6&     1.05&        0.42&        0.089&	150&81	\\         
7&     1.06&        0.31&        0.064&  	450&110 \\     
8&     0.60&        0.29&        0.047&	93&40	\\    
9&     0.51&        0.46&        0.051&	400&180	\\         
10&    0.59&        0.52&        0.064&	86&9.3	\\     
11&    0.62&        0.66&        0.070&	240&36	\\            
\hline
\end{tabular}
\end{center}
\label{tab:dataset}
\end{table}  

For a smooth model and a source close to the cusp, the flux 
relation described before holds. We see that the fluxes violate this 
rule in all configurations we used. As we mentioned before this
relation can only be violated when the source is away from the cusp or
if there is substructure in the system. Since here we know the source
position, the N-body lensing results show that the substructure 
we have in this particular simulated galaxy is indeed responsible for
the observed deviation.

\subsection{Discussion of the results from N-body lensing
\label{sc:dnbody}}
An important question is whether the N-body
simulation galaxy we are using is a good representation of a real
galaxy for the purpose of lensing. If the resolution is not high
enough, an N-body simulated galaxy might show more substructure than
a real galaxy has.

In order to obtain the surface mass density map representative of
lensing and to try to make sure that the substructure we see is not of
numerical origin we used a smoothing length for particles of
$\sigma=0.8 \: {\rm kpc} \sim 0.2 \:{\rm arcsec}$. The individual mass
clumps we see in the corresponding surface mass density map
(Fig.~\ref{fig:kappa_2k}) therefore contain well above 100 particles.
As we mentioned before, a realistic resolution scale for an identified
substructure in the simulation corresponds to $\sim 40$ particles.

The regions that are interesting for multiple image formation 
typically have $\kappa$ values of order unity. In such regions we have 
approximately 300 particles per smoothing area, which, if the noise were
Poissonian, would result in an error of about
$1/\sqrt{N}\sim 0.06$.  However, studies of errors of
N-body simulations using smoothed particle hydrodynamics  
\citep{mo92,ni78} show that the errors are significantly smaller, and
behave as  $\propto \log N / N$.

Any deviation of $\kappa$ due to substructure larger than the
estimated error of the N-body simulations tells
us that we are dealing with ``physical'' substructure,
i.e. not of numerical origin. Our surface
mass density maps show deviations well above the estimated Poisson
noise, and therefore also above numerical noise. 
 The results shown in Table~\ref{tab:dataset} are
affected by the ``physical'' substructure as well as by the
noise. Although the expected noise is significantly smaller
than believable substructure observed in the N-body simulations, 
it is hard to precisely estimate the relative contributions in the 
results shown 
here.

There are indications that the level of substructure as obtained from
simulations can
influence lensing phenomena a lot. In particular, the synthetic fluxes we
obtained deviate strongly from those predicted by smooth models. 
This particular example of a simulated galaxy can of course not
give us the answers to the aforementioned questions. In
order to draw stronger conclusions, one would have to investigate many
different realisations of N-body simulated galaxies and in addition
use higher resolution simulations (currently unavailable). A
statistical analysis to investigate the strong lensing properties 
could then be made. This is, however, beyond the scope of
this paper and will be considered in a future work.

\section{Conclusion \label{sc:concl}}
In this work we have investigated the influence of
substructure in the gravitationally lensed system B1422+231.  
While it is intuitively clear that a lens galaxy is not a smooth
entity, we have tried to investigate how deviation from a smooth model
can influence lensing phenomena, especially the image
flux ratios.

We have used two different smooth models for the lensing galaxy (SIE+SH
and NIE+SH), and both failed very badly in fitting the image
fluxes (we got $\chi^2=130$ with 2 degrees of freedom). The use of models 
with substructure requires additional observational
constraints. Therefore, we used deconvolved image
shapes as constraints. We get a significant improvement of the fit
compared to the smooth model. However, the way the substructure is
introduced is oversimplified, thus we should not be surprised that the
resulting $\chi^2$ is still high. For the model with a single
perturber we got $\chi^2=19$ for 6 degrees of freedom, and 
with two perturbers we had
$\chi^2=15$ for 4, while the model without
substructure (where deconvolved image shapes were included) gives
$\chi^2=140$ for 8 degrees of freedom.

Up to now we have not considered the possibility that microlensing
plays a role for the radio fluxes.  \citet{kop00} claim that they 
have detected microlensing in the multiply-imaged radio source B1600+434. 
Microlensing is a very tempting explanation for difficulties in
fitting the fluxes for it can also explain
why the $8.4 \: {\rm GHz}$ A:B flux ratio has changed from 0.97 in 1991 \citep{pa92} to 0.93
in 1997 \citep{pa99}.        
This has a consequence that again speaks in favour of
substructure, since the presence of radio microlensing indicates that
there is a significant number of compact objects in the lens galaxy halo.

N-body simulation data of a model galaxy provides a test for the 
influence of mass-substructure in strong gravitational lensing. When 
we generated data of four image
systems with the simulated galaxy we again experienced
difficulties. We have tried
to fit image positions and fluxes and failed to obtaining a model that
fits well. From these experiments we can conclude
that the level of substructure obtained from this particular N-body 
simulated galaxy can cause the same difficulties as experienced in some
of the real gravitationally lensed systems.

In order to obtain stronger conclusions 
one would have to investigate more
realisations of simulated galaxies, also at different
redshifts. However, the N-body
simulation results indicate that substructure plays an important
role in strong lensing. Also, modelling B1422+231 shows that
the fluxes of more than one image are probably affected by the lens
substructure. With the
conclusions one can draw from lens modelling we can not give a direct
answer about the properties of mass substructure in B1422+231.

One should
therefore avoid using the flux constraints directly; they should, rather,
be treated in statistical manner, e.g. in a way suggested by \citet{ma98}.   
Fortunately, the perturbations on the scales we
are dealing with here do not influence the image positions significantly,
and play even less of a role for the time delay \citep{ma98}. 
Strong lensing thus
remains one of the best tools to constrain the Hubble constant.

\begin{acknowledgements}
We would like to thank the referee for many constructive comments which have
helped to greatly improve this paper. Further we would like to thank 
Douglas Clowe and Alok Patnaik for many useful
discussions. This work was supported by the Bonn International Physics
Programme, by the Deutsche
Forschungsgemeinschaft, 
and by the TMR Network ``Gravitational Lensing:
New Constraints on Cosmology and the Distribution of Dark Matter'' of
the EC under contract No. ERBFMRX-CT97-0172.
\end{acknowledgements}

\bibliography{bibliogr} 
\bibliographystyle{aa}
\end{document}